\documentclass[numbers,sort&compress]{IntechOpen-Book}

\graphicspath{{Artworks/}}
\usepackage{listings}
\usepackage{caption}
\usepackage[frozencache=true,cachedir=.]{minted}

\usepackage{fancyhdr}
\def\makeheadbox{
						\hfill This chapter was accepted on May 12, 2021 for publication in Coding Theory - Recent Advances, New Perspectives and Applications, IntechOpen \hfill
					}
		
		\def\makeheadbox{{%
				\vspace{-1em}
				\hbox to0pt{\vbox{\baselineskip=10dd\hrule\hbox
						to\hsize{\vrule\kern3pt\vbox{\kern3pt
								\hbox{ \footnotesize \hspace{6em}This chapter was accepted on May 12, 2021 for publication in the book}
								\hbox{\footnotesize \hspace{2em} Coding Theory - Recent Advances, New Perspectives and Applications, IntechOpen, London}
								\kern3pt}\hfil\kern3pt\vrule}\hrule}%
					\hss}}}

\begin{document}

		\lstset{
		keywordstyle=\color{blue},
		breaklines=true,
		postbreak=\mbox{\textcolor{red}{$\hookrightarrow$}\space},
		columns=fullflexible,
		basicstyle=\ttfamily\footnotesize,
	}

\Mainmatter

\begin{frontmatter}

\chapter{Conversational Code Analysis: The Future of Secure Coding}
\author{Fitzroy D. Nembhard}
\author{Marco M. Carvalho}
\makeheadbox
\makechaptertitle

\chaptermark{Conversational Code Analysis: The Future of Secure Coding}

\begin{abstract} 
The area of software development and secure coding can benefit significantly from advancements in virtual assistants. Research has shown that many coders neglect security in favor of meeting deadlines. This shortcoming leaves systems vulnerable to attackers. While a plethora of tools are available for programmers to scan their code for vulnerabilities, finding the right tool can be challenging. It is therefore imperative to adopt measures to get programmers to utilize code analysis tools that will help them produce more secure code. This chapter looks at the limitations of existing approaches to secure coding and proposes a methodology that allows programmers to scan and fix vulnerabilities in program code by communicating with virtual assistants on their smart devices. With the ubiquitous move towards virtual assistants, it is important to design systems that are more reliant on voice than on standard point-and-click and keyboard-driven approaches. Consequently, we propose MyCodeAnalyzer, a Google Assistant app and code analysis framework, which was designed to interactively scan program code for vulnerabilities and flaws using voice commands during development. We describe the proposed methodology, implement a prototype, test it on a vulnerable project and present our results.
\end{abstract}

\begin{keywords} 
secure coding, virtual assistant, code analysis, static analysis
\end{keywords}

\end{frontmatter}

\section{Introduction}

Computing systems face serious threats from attackers on a day-to-day basis. Devices within a network could be targeted or used as launching pads to spawn malware and other attacks to critical systems and infrastructure. A system is as secure as its weakest link \cite{schneier16secrets}. 	Therefore, software engineers must be cognizant of the cyber-related challenges that plague modern computer systems and engineer software with credible defenses. One of the first defenses against potential threats to computer systems is careful analysis of program code during development and taking necessary steps to minimize/eliminate vulnerabilities. 

Program analysis falls into three main categories:  static application security testing (SAST) or static analysis, dynamic application security testing (DAST) or dynamic analysis, and interactive application security testing (IAST). Static	
analysis is a \textquotedblleft technique	in	which	code	listings,	 test	results,	or	other	documentation are \ldots examined \ldots to	identify errors,	violations	of	development	standards,	or	other	problems\textquotedblright \cite{ieee_vocab}. Dynamic analysis is the \textquotedblleft process	of	evaluating	a	system	or	component	based	on	its	behavior	during	execution\textquotedblright \cite{ieee_vocab}. IAST involves instrumenting a program with sensors to monitor program code in memory during execution in order to find specific events that could cause vulnerabilities \cite{nembhard_hybrid}. Two or more of these approaches may be combined to create hybrid tools and techniques for analyzing program code.  These hybrid systems are designed to achieve more comprehensive coverage and to decrease the false positives and false negatives of existing approaches.

 While researchers are interested in designing sound and complete code analysis tools,  achieving soundness and completeness remains an intractable problem \cite{Heckman_systematic_asa_lit_review, static_analysis_for_security, chess2007secure}. Consequently, a lot of research in code analysis is centered on improving the alerts of static analysis tools \cite{muske_aait_2016_survey, Heckman_systematic_asa_lit_review}. More recently, several researchers have proposed models based on deep learning and other machine learning approaches to scan and fix vulnerabilities in program code \cite{nembhard_towards_recommender}. Many of these tools are still at an infant stage and have not yet made it to market. Based on current trends, we believe that the future of code analysis will involve more refined tools based on artificial intelligence (AI), machine learning, and other hybrid approaches.
 
In this work, we propose a hybrid code analysis framework that employs the use of voice assistants (VAs) to allow a programmer to conversationally scan for and fix potential vulnerabilities in program code. The use of  voice assistants have grown significantly in recent years. This work focuses primarily on the Google Assistant\footnote{Google, Google Assistant, and Dialogflow are registered trademarks of Google, Inc. The use of these names or tools and their respective logos are for research purposes and does not connote endorsement of this research by Google, Inc. or any of its partners.} as it is the most popular \cite{klein2020exploring_vas} among other virtual assistants. 

The rest of the chapter is organized as follows: first, we discuss related work in the area of hybrid analysis in Section \ref{sec:related_work} followed by a discussion on challenges affecting adoption of existing approaches in Section \ref{sec:adoption_challenges}.  In Section \ref{sec:future_code_analysis}, we theorize about the future of secure coding and propose a new code analysis approach in Section \ref{sec:proposed_approach}. We then use a case study to evaluate our proposed approach in Section \ref{sec:case_study} and present our conclusion in Section \ref{sec:conclusion}.

\section{Related Work}
\label{sec:related_work}
This work falls in the area of hybrid analysis. In this section, we summarize works in this area. 

In \citeyear{Aggarwal_integrating_static_dynamic}, \citeauthor{Aggarwal_integrating_static_dynamic}  
combined static and dynamic analysis to detect buffer overflow in C programs. Both static and dynamic approaches have advantages and disadvantages. One of the disadvantages of dynamic analysis is the requirement of a large number of test cases, which present an overhead. Some dynamic analysis tools use a feature know as generate-and-patch or generate-and-validate in an effort to auto-fix vulnerabilities. In \citeyear{qi2015_kali_patching_tool_with_comparisons}, the authors of  \cite{qi2015_kali_patching_tool_with_comparisons} analyzed reported patches for several DAST tools including GenProg, RSRepair, and AE, and found that the overwhelming majority of reported patches did not produce correct outputs.  The authors attributed the poor performance of these tools to weak proxies (bad acceptance tests), poor search spaces that do not contain correct patches, and random genetic search that does not have a smooth gradient for the genetic search to traverse to find a solution \cite{qi2015_kali_patching_tool_with_comparisons}.

In \citeyear{Chebaro_2012_program_slicing_sast_dast}, \cite{Chebaro_2012_program_slicing_sast_dast} proposed a hybrid approach that uses source code program slicing to reduce the size of C programs while performing analysis and test generation. The authors used a minimal slicing-induced cover and alarm dependencies to diminish the costly calls of dynamic analysis \cite{nembhard2018_dissertation_recommender}.

In \citeyear{tripp2014hybrid}, \cite{tripp2014hybrid} implemented a hybrid architecture as the JSA analysis tool, which is integrated into the IBM AppScan Standard Edition product. The authors augmented static analysis with (semi-)concrete information by applying partial evaluation to JavaScript functions according to dynamic data recorded by the Web crawler. The dynamic component rewrites the program per the enclosing HTML environment, and the static component then explores all possible behaviors of the partially evaluated program.

In \citeyear{Kiss2015_program_slicing_sast_dast_heartbleed}, \cite{Kiss2015_program_slicing_sast_dast_heartbleed} applied a program slicing technique, similar to \cite{Chebaro_2012_program_slicing_sast_dast}, to create a tool called \textit{Flinder-SCA}. The authors also implemented their program using the \textit{Frama-C} platform. The main difference between \cite{Chebaro_2012_program_slicing_sast_dast} and \cite{Kiss2015_program_slicing_sast_dast_heartbleed} is that \cite{Kiss2015_program_slicing_sast_dast_heartbleed} performs abstract interpretation and taint analysis via a fuzzing technique wheres \cite{Chebaro_2012_program_slicing_sast_dast} does not perform taint analysis or fuzzing.

Also, in \citeyear{li2015hybrid}, \cite{li2015hybrid} proposed a hybrid malicious code detection scheme that was designed using an AutoEncoder and deep belief networks (DBN). The AutoEncoder deep learning method was used to reduce the dimensionality of data. The DBN was composed of multilayer restricted boltzmann machines (RBM) and a layer of BP neural network. The model was tested on the KDDCUP'99 dataset but not on actual program code.

In \citeyear{sapfix_facebook_2019}, \cite{sapfix_facebook_2019} proposed SapFix, a static and dynamic analysis tool which combines a mutation-based technique, augmented by patterns inferred from previous human fixes, with a reversion-as-last resort strategy for fixing high-firing crashes. This tool is built upon Infer \cite{infer_tool_facebook_2015} and a localization infrastructure that aids developers in reviewing and fixing errors rapidly. Currently, SapFix is targeted at null pointer exception (NPE) crashes, but has achieved much success at Facebook \cite{infer_tool_facebook_2015}.

In a dissertation produced in \citeyear{bhagyanath2021_hybrid_dissertation}, \cite{bhagyanath2021_hybrid_dissertation} proposed a code generation technique for  synchronous control asynchronous dataflow (SCAD) processors based on a hybrid control-flow dataflow execution paradigm. The model is inspired by classical queue machines that completely eliminates the use of registers. The author uses satisfiability (SAT) solvers to aid in the code generation process \cite{bhagyanath2021_hybrid_dissertation}.

To the best of our knowledge, our work is the first to employ modern virtual assistants to conversationally scan and fix vulnerabilities in program code. In \cite{austerjost2018_lab_vas}, the authors established a voice user interface (VUI) for controlling laboratory devices and reading out specific device data. The results of their experiments produced benchmarks of established infrastructure and showed a high mean accuracy (95\% ± 3.62) of speech command recognition and reveals high potential for future applications of a VUI within laboratories \cite{austerjost2018_lab_vas}. In like manner, we propose the integration of personal assistants with code analysis systems to encourage programmers to produce more secure code.
 
\section{Challenges Affecting Adoption of Existing Approaches}
\label{sec:adoption_challenges}
\begin{figure}[!h]\centering
	\FIG{\includegraphics[width=1.0\textwidth]{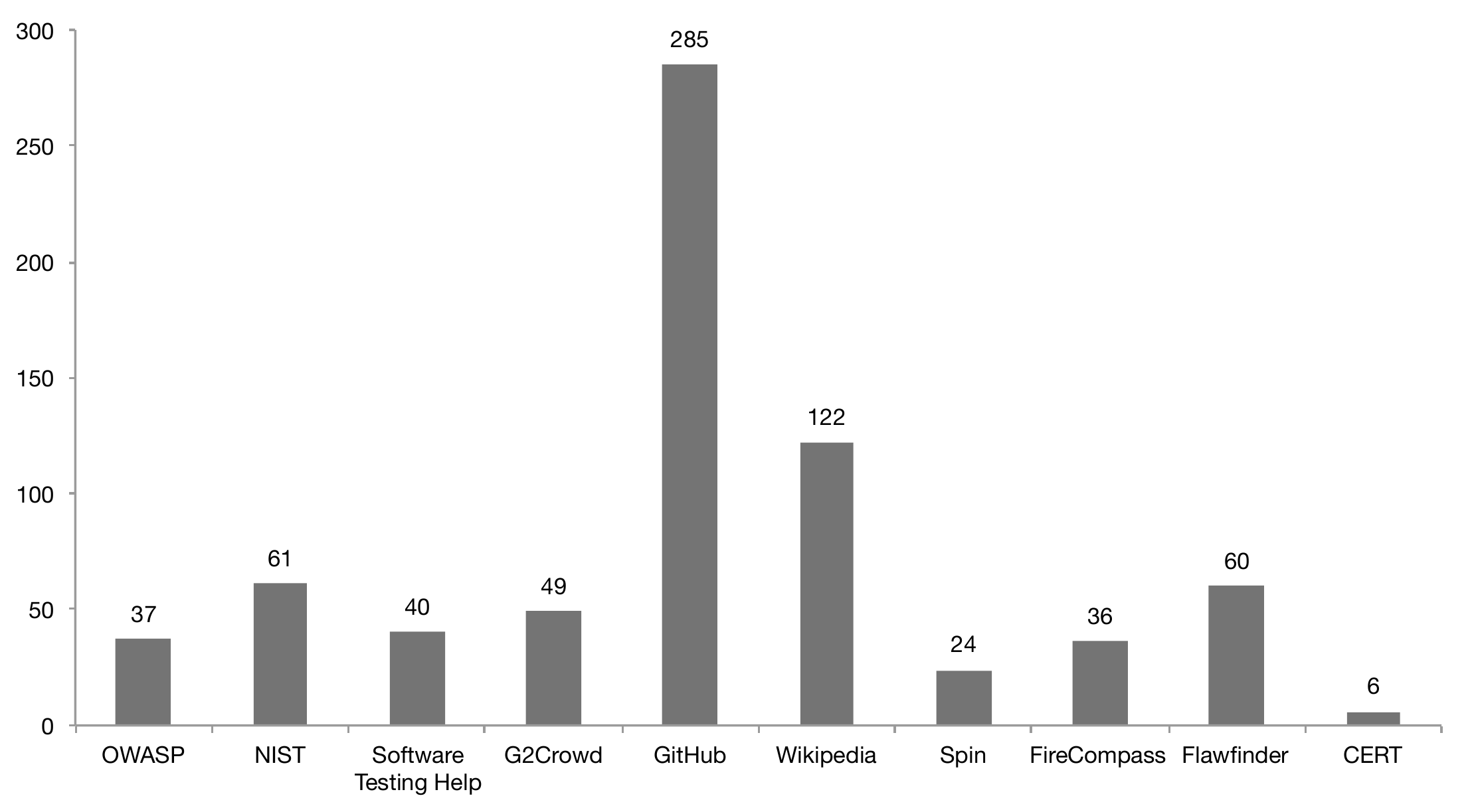}}
	{\caption{The large number of code analysis tools found on popular websites\label{fig:tools_barchart}}}
\end{figure}

\begin{figure}[!h]\centering
	\FIG{\includegraphics[width=1.0\textwidth]{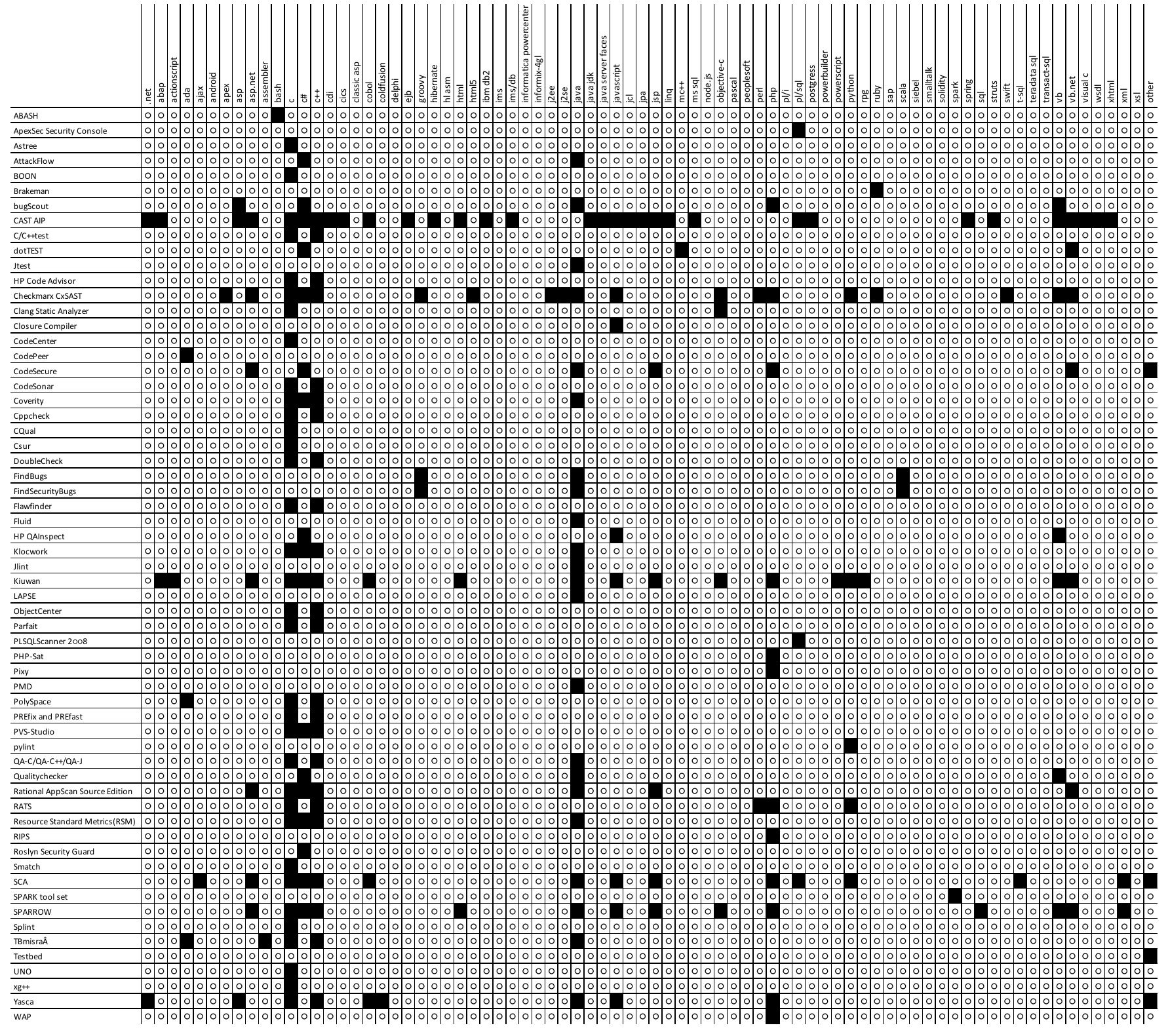}}
	{\caption{Static analysis tools categorized by programming language\label{fig:tools_by_language}}}
\end{figure}

Several code analysis and vulnerability detection surveys have categorized tools in the literature \cite{muske_aait_2016_survey, 2016_static_analysis_alarms_survey, 2015_static_analysis_tools_tech, vulnerability_detection_2020_survey}. While surveys are essential in advancing research, many of them do not focus on tools found on websites. It must be noted that the average programmer does not look for tools in research papers. To that end, we conducted a Google search and found several popular websites that present various tools that programmers may use to scan their code for vulnerabilities. Figure \ref{fig:tools_barchart} shows a bar chart highlighting the number of tools found on these websites. As shown in the figure, GitHub and Wikipedia list the most tools and are often the top websites returned in search results due to their popularity. We further grouped the most popular static analysis tools found on these websites by language as shown in Figure \ref{fig:tools_by_language}. As can be seen, this non-exhaustive list could overwhelm many programmers in determining the best tools for their projects. 

In addition, the ability to combine code analysis approaches coupled with the number of programming languages that exist result in a large number of tools from which coders can choose to analyze their code. This makes it onerous for a programmer or organization to decide on a particular code analysis tool. Further, tools often require special configuration, which may take time to fine tune for best results. Many tools also suffer from usability issues, lengthy vulnerability reports, and false positives, making programmers avoid them altogether \cite{nembhard_interface_impact, developers_dont_use_sast_tools, kremenek_correlation_2004}.  

Another challenge affecting adoption of code analysis tools is monopolization of the market by certain companies. For-profit companies usually have the resources to improve tools by adding more state-of-the-art approaches such as cloud-based scanning, IAST support, and report generation. While these developments often advance the field of code analysis, they sometimes discourage small organizations and individuals from investing the effort and resources required to procure state-of-the-art tools.  Thus, a streamlined, modern, cost-effective approach is needed to help encourage programmers to produce more secure code. 

\section{The Future of Code Analysis}
\label{sec:future_code_analysis}
We believe that the future of code analysis lies in hybrid systems that combine several approaches to achieve useful analyses and actionable reports that will encourage programmers to produce more secure software. Based on current trends in machine learning, especially in deep learning, and natural language processing (NLP) (e.g., virtual assistants), it is safe to say that future code analysis will rely heavily on AI, ontologies, NLP, and machine learning. For example, when discussing the trends and challenges of machine learning, the authors in \cite{trends_ml_2018} \textquotedblleft envision a fruitful marriage between classic logical approaches (ontologies) with statistical approaches which may lead to context-adaptive systems (stochastic ontologies) that might work similar to the human brain\textquotedblright \cite{trends_ml_2018}.

\begin{figure}[!h]\centering
	\FIG{\includegraphics[width=0.9\textwidth]{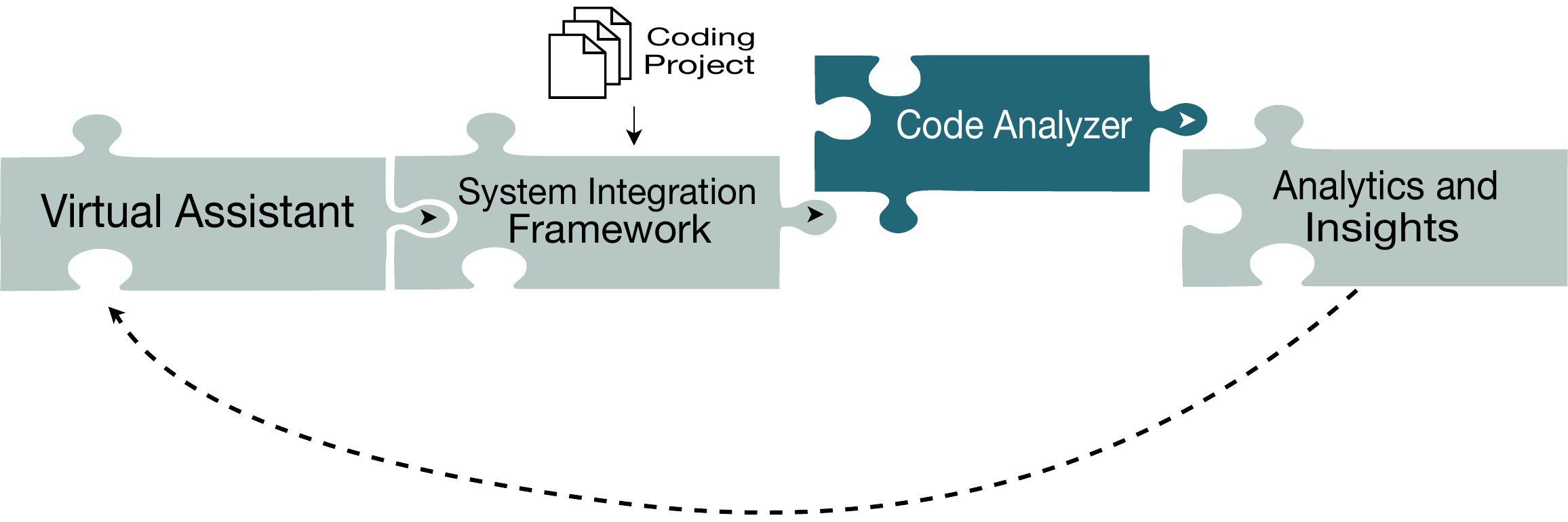}}
	{\caption{A suggested model showing code analysis as part of a plug-and-play paradigm that facilitates the inclusion of any analysis tool and the use of a virtual assistant to manage the analysis process.\label{fig:pnp_model}}}
\end{figure}

Our projection is that code analysis frameworks will facilitate plug-and-play (PnP) models. Figure \ref{fig:pnp_model} illustrates a generalized PnP model that uses virtual assistants to manage the analysis process. Using this plug-and-play model, programmers may select the code analyzer that best fits their project based on factors such as project type, project size, speed, efficiency, security, etc. 
This is similar to the current landscape with virtual assistants and recommender systems. Currently, a person may use a virtual assistant like the Google Assistant to navigate a list of restaurants based on price, location, menu, reviews, etc. The virtual assistant may update the users preferences based on selections over time. This concept can also apply in code analysis where the chosen scanner used in the PnP model could be based on past scans or popularity.

The code analyzer featured in the model in Figure  \ref{fig:pnp_model} may use any combination of approaches including SAST, DAST, and IAST, which could be cloud-based or localized to the user's computer. These  approaches could be backed by any algorithm that results in significant performance gains. It has been shown in the literature that deep learning and other ensemble methods perform very well in a large number of contexts including infected host detection \cite{infected_hosts_2020}, intrusion detection systems \cite{improving_ids_ensemble_2018, deep_learning_ids_2017}, and malware analysis \cite{deep_learning_malware_2016, ml_for_malware_2020}, to name a few. Interestingly, many of these approaches can be used to create or improve code analyzers in an effort to help programmers produce more secure software.  

\begin{figure}[!h]\centering
	\FIG{\includegraphics[width=1.0\textwidth]{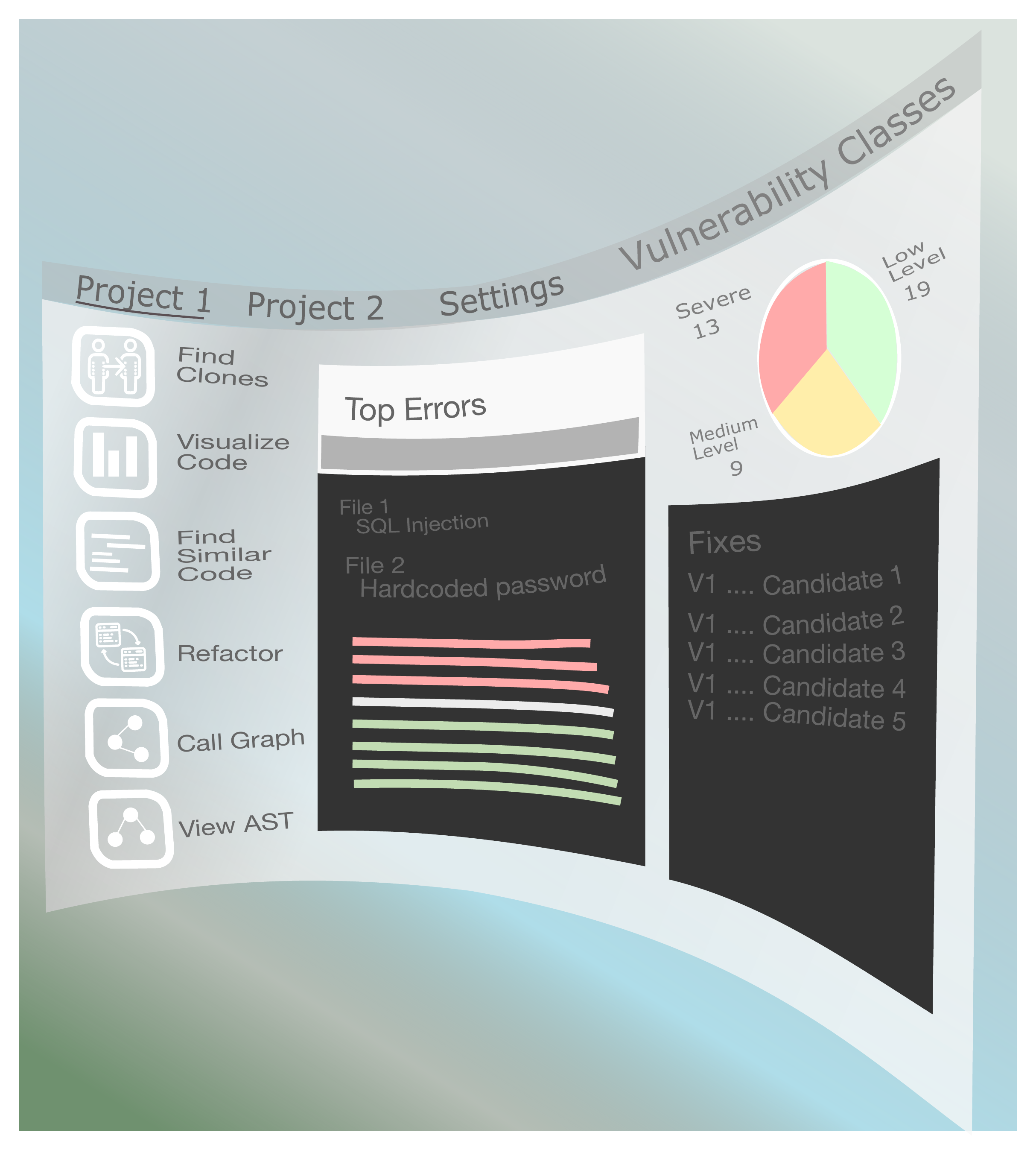}}
	{\caption{A mockup of an analytical dashboard for code analysis on a curved display\label{fig:ui_mockup}}}
\end{figure}

Another feature of code analyzers of the future is a deep reliance on data analytics, visualizations and state-of-the-art interfaces. As discussed in the literature \cite{nembhard_interface_impact, nembhard_towards_recommender}, the interface of a code analyzer can have a negative or positive impact on its use and adoption. Therefore, for a system to be adopted in any project or organization, users must be able to gain insights from the way it presents its results. Figure \ref{fig:ui_mockup} shows a mockup of what we believe the interface of future code analyzers will look like. These interfaces will be in the form of dashboards instead of the customary lengthy bug reports displayed in a console. 

\section{Proposed Approach}
\label{sec:proposed_approach}
The proposed approach is to integrate a virtual assistant  with a code analysis framework that allows users to scan, analyze, refactor and fix their code of inconsistencies and vulnerabilities. In this section, we describe the proposed approach using the system architecture.
 
\subsection{System Architecture}
\label{sec:system_architecture}
\begin{figure}[!h]
	\begin{center}
		\includegraphics[width=1.0\textwidth]{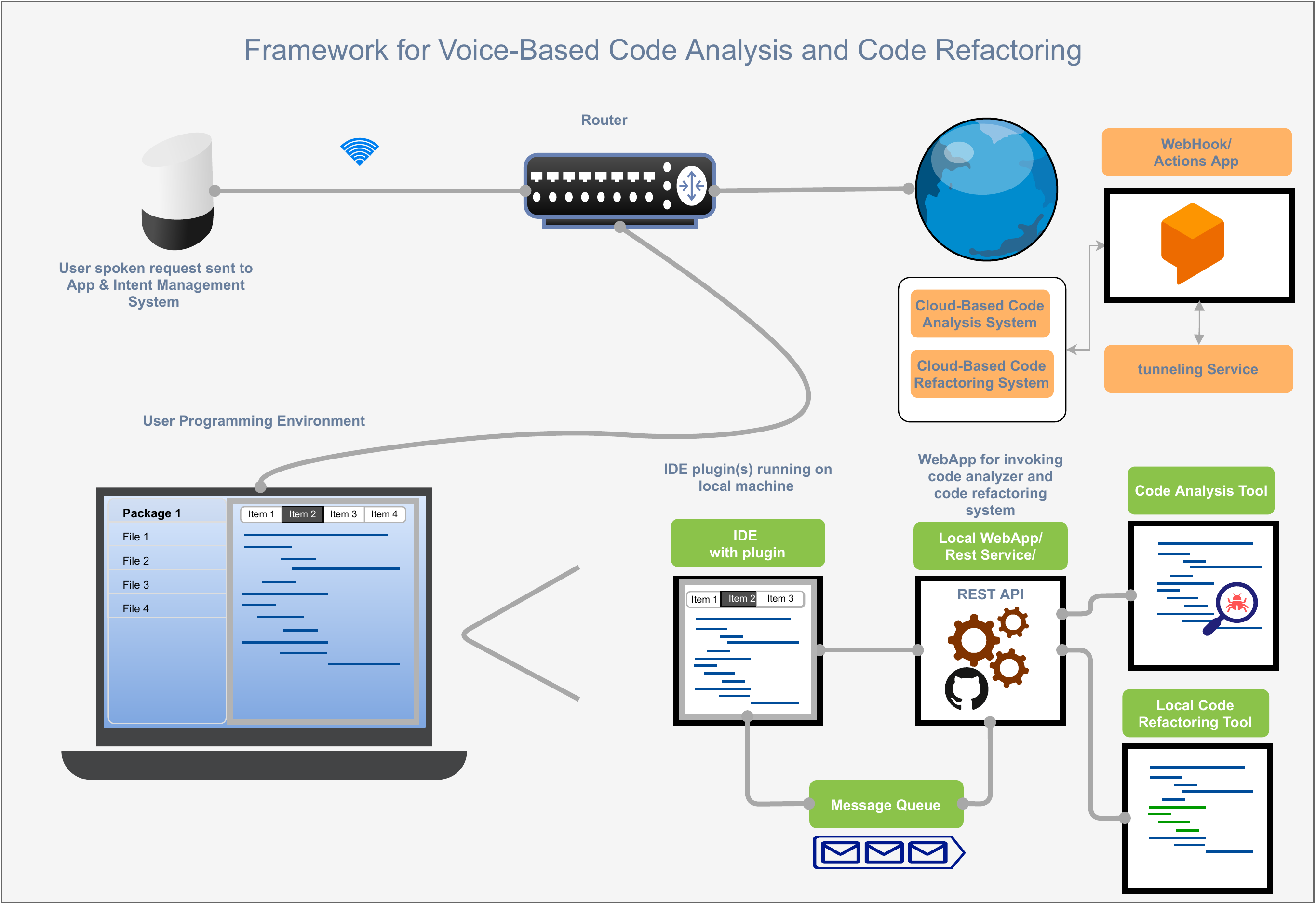}
		\caption{MyCodeAnalyzer system architecture}
		\label{fig:system_architecture}
	\end{center}
\end{figure}

\begin{figure}[!h]\centering
	\FIG{\includegraphics[width=1.0\textwidth]{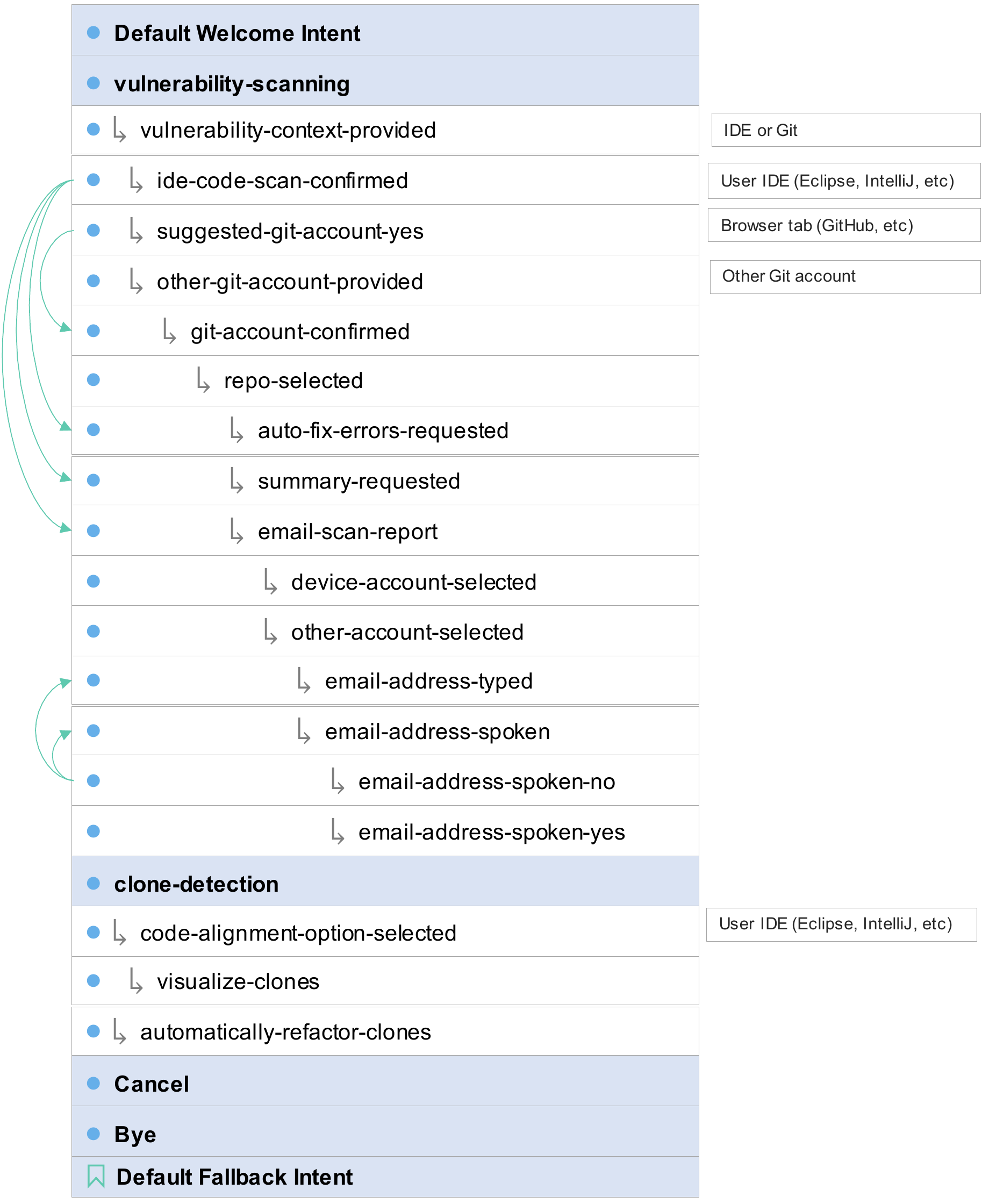}}
	{\caption{Current Dialogflow intents used by MyCodeAnalyzer \label{fig: dialogflow_intents}}}
\end{figure}

The system architecture for MyCodeAnalyzer is shown in Figure \ref{fig:system_architecture}. The system consists of three main components: the virtual assistant, the webhook API and the code scanning environment. The code scanning environment consists of a web app, an integrated development environment (IDE) plugin, code analyzers and refactoring tools. Google Assistant was chosen as the virtual assistant  because of its popularity and easy-to-use App Engine and Dialogflow frameworks. The process flow is as follows: a user invokes a Google Assistant app (aka, Google Actions app) using a set of phrases understood by the system. 
This app is specially designed to understand trigger phrases associated with code analysis. Trigger phrases are training phrases that are entered into Dialogflow using an intent management system.  Dialogflow is a natural language understanding platform that allows users to design and integrate a conversational user interface into a mobile app, web application, device, bot, interactive voice response system, etc. \cite{dialogflow}. Figure \ref{fig: dialogflow_intents} captures the current intents incorporated into MyCodeAnalyzer. Each intent is backed by machine learning and NLP technology that uses named entity recognition (NER) and other approaches to extract entities from speech, determine context, and carry out tasks.  

The intents in MyCodeAnalyzer are organized into 6 main categories: \textit{Default Welcome Intent, vulnerability-scanning, clone-detection, Cancel, Bye, and Default Fallback Intent}. The \textit{Default Welcome Intent} is used to welcome the user to the system and provide a description of potential requests that the application can fulfill. The \textit{vulnerability-scanning} intent is the most complex of the intents and uses a tree-like structure to allow the user to conversationally scan a project for vulnerabilities, email a scan report or auto-fix errors based on the capabilities of the code analyzer. The \textit{clone-detection} intent is used to scan a project for duplicated code and to provide a visualization showing a side-by-side comparison of similar code. While clones may not be vulnerable, they could become bloat in a project and could potentially lead to vulnerabilities. The \textit{Cancel} intent is used to exit a task currently underway. \textit{Bye} is used to exit the system and the \textit{Default Fallback Intent}, as the name suggests, is used to ask the user to repeat a phrase for clarification or serve as a graceful fail mechanism.

\begin{figure}[!h]
	\begin{center}
		\includegraphics[width=1.0\textwidth]{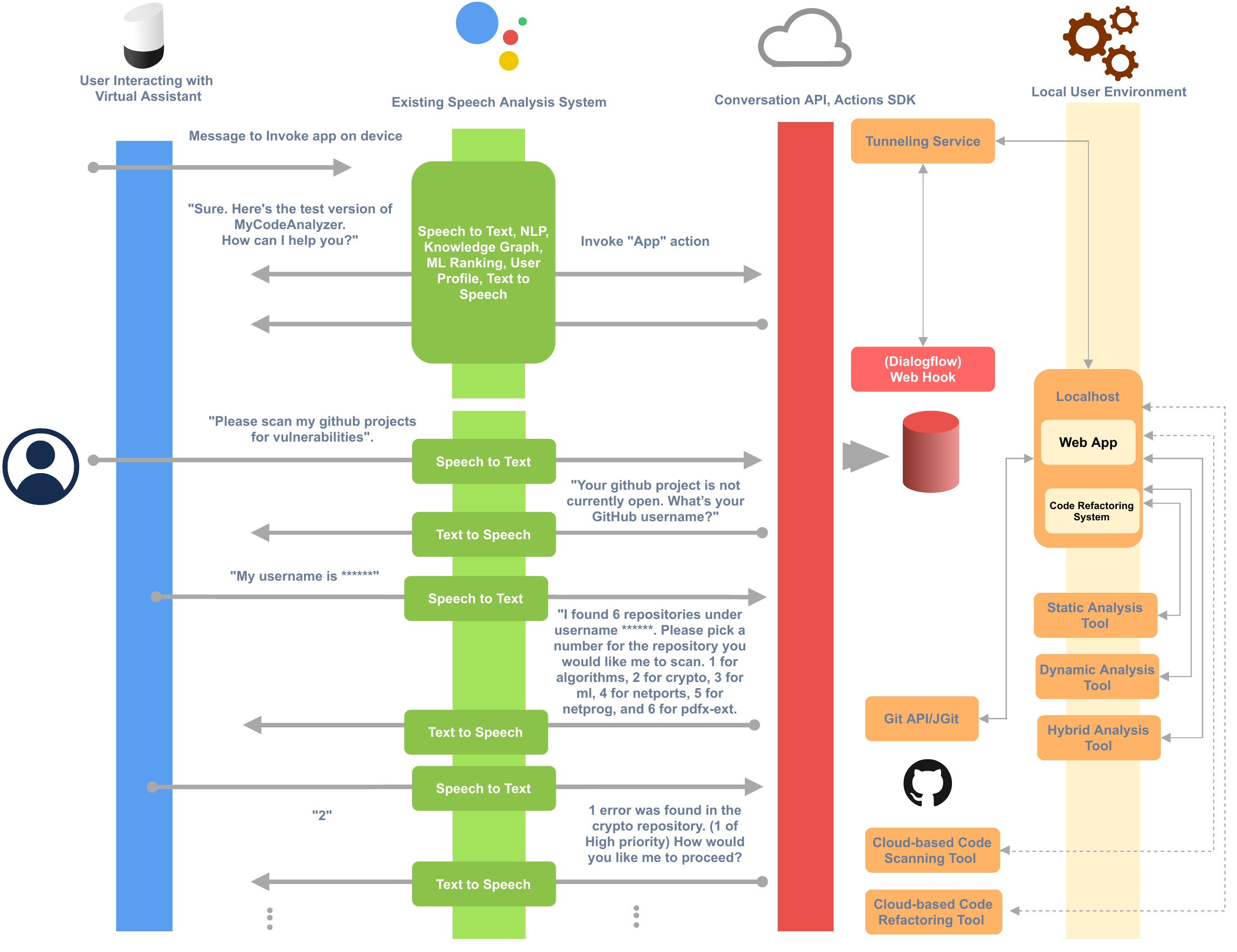}
		\caption{Internals of MyCodeAnalyzer showing the flow of information throughout the system}
		\label{fig:system_internals}
	\end{center}
\end{figure}

Once invoked, the Google Assistant app communicates with the Google Conversation API to determine the user's intent. After intent has been determined, the Google Actions app then uses webhooks to communicate with a web service running on the user's computer. Using a tunneling service, the web service interacts with the user's IDE by way of a plugin. This plugin invokes a code analyzer or refactoring tool, takes actions based on the user's request, and places a message in a message queue. The web service then reads the queue and returns the message to the Google Assistant app, which then reads the message back to the user.  The webhooks were set up in Dialogflow and run as servlets on  Google App Engine. A servlet accepts valid Dialogflow POST requests and responds with data that is processed by the Google Assistant app and returned as output messages to the user.  Figure \ref{fig:system_internals} further shows the internals of the system during a conversation between the user and the assistant. While only the static analysis portion of the system is demonstrated in this work, the system is modular enough for dynamic and hybrid analysis tools to be incorporated using the PnP approach discussed in Section \ref{sec:future_code_analysis}. This approach provides a more complete code analysis depending on the user's preferences.

\subsection{Accessing Information About The Coding Environment}
Two types of code-related information are accessed on the user\textquoteright s computer: code within the IDE and code from a Git repository (e.g., GitHub) currently opened in a web browser. The first type of information is important because it helps us to scan code being actively developed, while the second type is used in the case where the user would like to ensure that a repository is safe before forking it. MyCodeAnalyzer can detect GitHub pages that are open in a browser. On systems running MacOS, Applescript is used to communicate with the web browser.  Other approaches will be employed in the future to reproduce this functionality on machines running other operating systems.

In order to access the user\textquoteright s computer to scan the code being worked on in the IDE or referenced in the browser, a methodology must be established to access this information in a minimally invasive manner. To do so, we created a plugin for a given IDE. Currently, we have plugins for IntelliJ IDEA and Eclipse. The plugin becomes a part of the IDE, monitors the code being developed, and updates a message queue (data file) with information about the code files and projects manipulated by the programmer. Also, special system calls are used to access any browser tabs that point to GitHub projects.  A local web app in the form of a Spring MVC REST API \cite{spring_boot} runs on the user\textquoteright s computer. The job of the local web app is to communicate with MyCodeAnalyzer by way of a tunnel in order to scan local code or GitHub projects displayed in the user\textquoteright s  web browser.

\subsubsection{Accessing Code within the IDE}
Listing \ref{lst:applescript_all_open_apps} shows the Applescript code that is used to check for gui-based applications that are currently open on the user\textquoteright s computer. Following this is a snapshot of the corresponding output, which includes the Intellij IDEA IDE in the list. This Applescript code is added to the REST app where it is run on localhost and invoked by MyCodeAnalyzer to determine if the user is actively using an IDE. To further contextualize the process of determining which code the user would like to scan, it is also of interest to find out the \textit{frontmost} or most active application on the user's computer. To do so, the code shown in Listing \ref{lst:applescript_for_most_active_app} was used. This code is expected to return a single application, which in turn allows MyCodeAnalyzer to return a more direct response to the user. For example, a response might be, \textit{"Say IDE, if you would like me to scan the code that you are currently working on in IntelliJ"} instead of using indirect phrases such as \textit{"...may be working on."} 

\begin{listing}[h]
\begin{minted}[frame=lines]{Applescript}
set text item delimeters to ", "
tell application "System Events" to 
(name of every process where background only is false) as text
end tell
\end{minted}
\caption{Applescript code used to list all gui-based applications that are currently running on the user\textquoteright s computer }
\label{lst:applescript_all_open_apps}
\end{listing}

\noindent The following is a sample output generated using the code in Listing \ref{lst:applescript_all_open_apps}:
\begin{verbatim}
"Google Chrome, Sublime Text, Terminal, idea, pycharm,Teams,
Mail, teXShop, Notes, Spotify, Finder, Microsoft PowerPoint,
X11.bin, AdobeReader, iTunes, Microsoft Excel, Script Editor,
Activity Monitor, System Preferences, Safari, Preview"
\end{verbatim}

Since most IDEs are standalone applications, we believe the best way to have access to the user\textquoteright s code in a minimally invasive manner is to be an \textquotedblleft insider\textquotedblright\ (That is, to use a plugin that becomes part of the IDE).  Consequently, the goal of the plugins was to monitor the code being developed by taking note of the coding project and the coding files being manipulated by the user. To accomplish this, listeners were added to the IDE to detect when the text editor portion of the IDE is active, when tabs are activated or switched, and when code files are edited. The message queue is updated with the following pieces of information when the aforementioned actions are performed: \textit{ProjectName, ProjectLocation, CurrentFile, DateAdded, CurrentlyActive}. This queue is then queried for active files and projects when POST requests are made by the Google Assistant app to the local REST service running on the user's computer.

\begin{listing}[h]
\begin{minted}[frame=lines]{Applescript}
tell application "System Events"
name of application processes whose frontmost is true
end tell
\end{minted}
\caption{Applescript code used to determine the most active application on a computer}
\label{lst:applescript_for_most_active_app}

\end{listing}

\subsubsection{Accessing Code Referenced by Tabs Opened in the Web Browser}
Like IDEs, web browsers  provide little to no way for outside tools to access their core areas. However, the Applescript-based techniques used previously for accessing the System Events utility can be used to access the tabs that are currently open in the web browser on the user\textquoteright s device. Listing \ref{lst:applescript_for_most_browser_tabs} is used to retrieve tabs currently open in Google Chrome. This script can be modified to get tabs in other browsers such as FireFox or Safari.
MyCodeAnalyzer then checks if any of the URLs point to valid public GitHub accounts, which are then searched for coding projects if the user requests that a scan of a Git project be performed.

\begin{listing}[h]
\begin{minted}[frame=lines]{Applescript}
set text item delimeters to ","
tell application "Google Chrome" to URL  of tabs of every window 
as text
end tell
\end{minted}
\caption{Applescript code used to retrieve tabs currently open in Google Chrome. }
\label{lst:applescript_for_most_browser_tabs}
\end{listing}

\section{Case Study}
\label{sec:case_study}
In this section, we present a case study that demonstrates an implementation of our proposed methodology. The main goal of this case study is to demonstrate the applicability of integrating a virtual assistant into a code analysis framework to allow the user to conversationally scan their code for vulnerabilities. The system is currently in a prototypical stage. Here we perform a scan of a coding project using the Google Assistant app via an Apple iPhone.

 The following was done based on the proposed approach discussed in Section \ref{sec:proposed_approach}:
\begin{enumerate}
	\item Create a Google Assistant app\\
	A Google Assistant app was created based on the intents depicted in Figure \ref{fig:system_internals}. Dialogflow, Google App Engine, and Google Actions Console are key components in the design of the app. Once designed, the app was tested using the Google Actions API Simulator as well as released in alpha mode and tested on a smart phone running the Google Assistant. 
	
	\item Create a local web app to interface with the Google Assistant app and the coding environment\\
	The local web app was created using Spring Boot \cite{spring_boot} and was launched on the computer via Apache Tomcat \cite{apache_tomcat}.
	
	\item Create an IDE plugin for IntelliJ IDEA\\
	Our IntelliJ IDEA plugin was created and installed in IntelliJ version 2020.3.2. The plugin was installed using the IntelliJ plugin installer, which installs a local plugin from a JAR (Java ARchive) file.  
	
	\item Choose and integrate a code analyzer\\
	PMD \cite{pmd_static_analysis} static code analyzer (version 6.31.0) was chosen for this study. PMD uses a rule-based system to find common programming flaws in code written in 8 programming languages, offering the most support for Java and Apex. The rules used by PMD are divided into categories such as best practices, error prone, and security. For this case study, a set of rules was selected from the error prone and security categories.
	
	\item Chose a vulnerable project\\
	The OWASP WebGoat \cite{owasp_webgoat} project was used to evaluate the system. WebGoat is an insecure application that allows researchers and developers to test vulnerabilities commonly found in Java-based applications that use common and popular open source components \cite{owasp_webgoat}. 
	
	\item Test the system and report results \\
	To integrate the Google Actions app with the local web app, Ngrok \cite{ngrok} was chosen as the tunneling tool. Ngrok is a tool that exposes local servers behind NATs and firewalls to the public Internet over secure tunnels \cite{ngrok}.
	
\end{enumerate}

\subsection{Results and Discussion}
In this section, we capture a conversation between the Google Assistant app during the analysis of the WebGoat Project, present the report generated by the assistant, and discuss the results.
It must be noted that the errors found by the Assistant during the code analysis are the same as those that would be produced by the standalone PMD project.

\begin{figure}[!h]
	\begin{center}
		\includegraphics[width=0.55\textwidth]{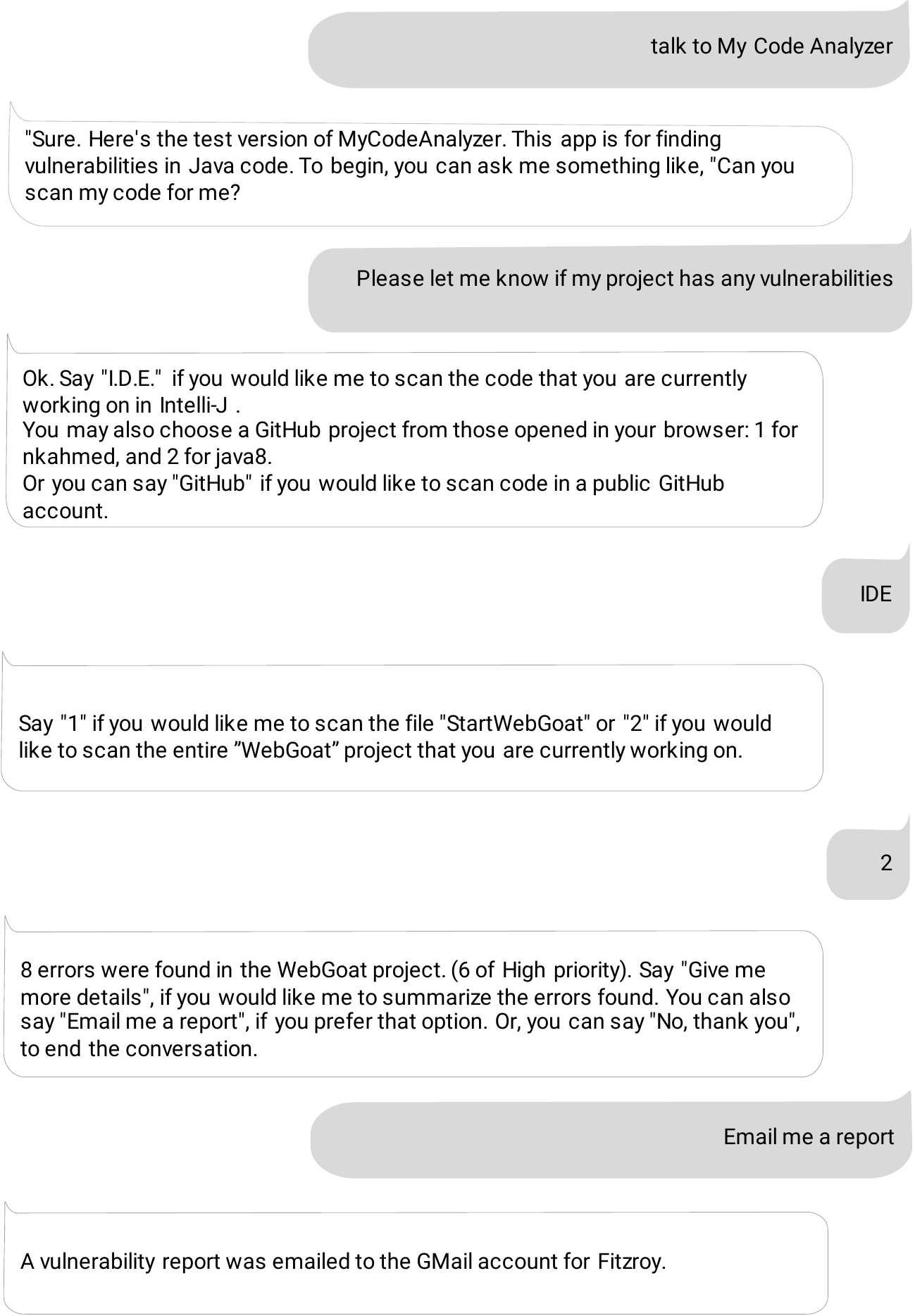}
		\caption{A conversation between MyCodeAnalyzer and a human tester while scanning the OWASP WebGoat project }
		\label{fig:analysis_conversation}
	\end{center}
\end{figure}

\begin{figure}[!h]
	\begin{center}
		\includegraphics[width=1.0\textwidth]{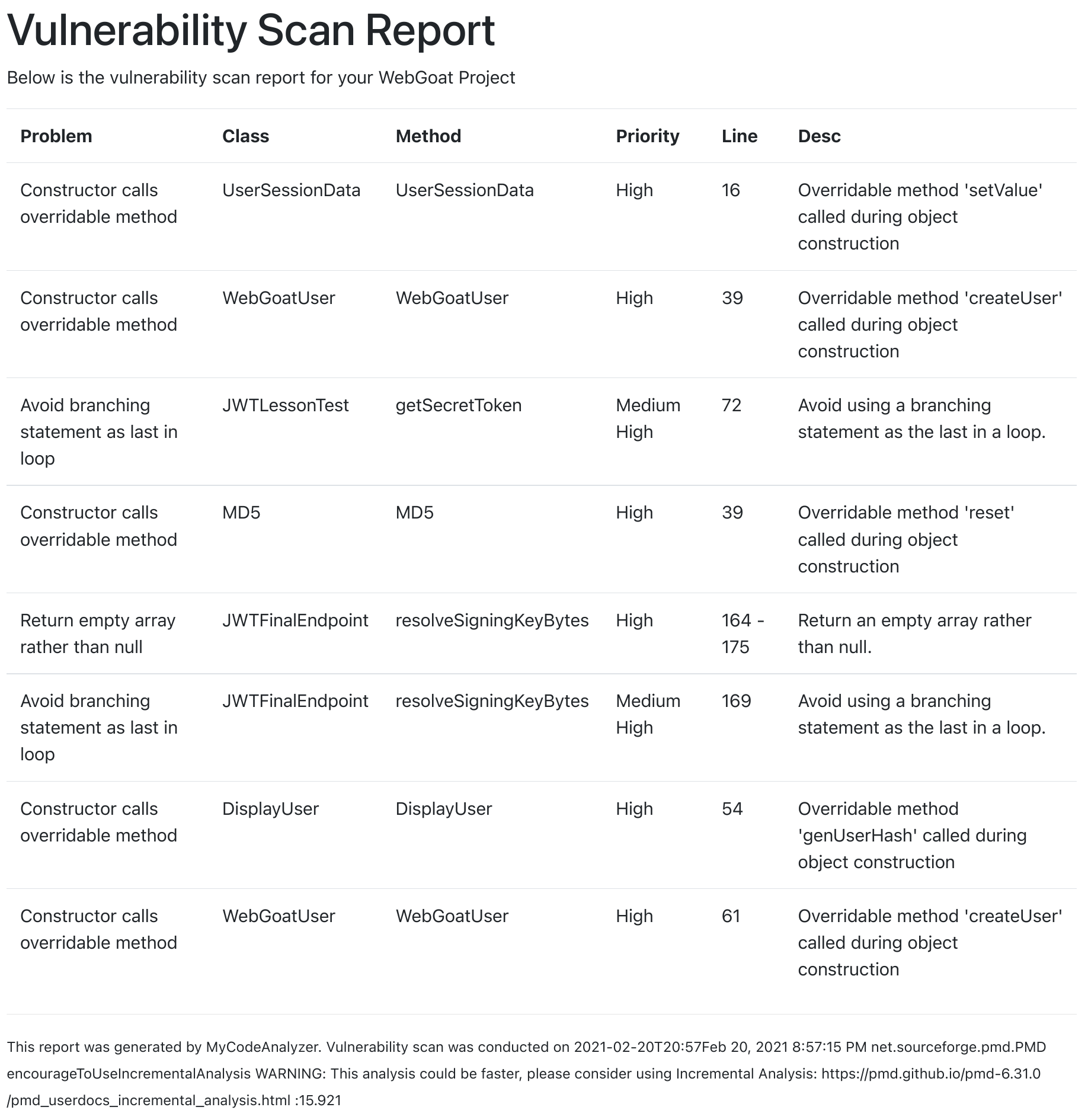}
		\caption{The report generated by MyCodeAnalyzer and emailed to the user after scanning the OWASP WebGoat project }
		\label{fig:vulnerability_report}
	\end{center}
\end{figure}

At this early stage of the project, the main benefit of the system is the ability to use a virtual assistant to perform code analysis while multitasking, thus improving productivity. After the system is setup, the programmer can configure and engage with the VA by voice without having to manually configure the code analyzer or browse and try to understand lengthy bug reports. The assistant can be used to perform actions based on the severity of the vulnerabilities found in the project. In the current version of MyCodeAnalyzer, Google Assistant can email the user a well-formatted report or read out the most important action items after analyzing the code. Figure \ref{fig:analysis_conversation} captures a conversation between a human tester and the Google Assistant. Figure \ref{fig:vulnerability_report} shows a formatted vulnerability report generated by the assistant and emailed to the user after scanning the WebGoat project. The WebGoat project has more severe vulnerabilities, but only those in the figure were captured by PMD based on  the rulesets used by the analyzer. As can be seen from the report, MyCodeAnalyzer was able to process the lengthy XML report returned by PMD into a more easily understood report that captures only pertinent information. These results demonstrate the applicability of using a framework backed by virtual assistants to scan code for vulnerabilities and generate meaningful reports.

\subsection{Challenges}
It is important to outline some challenges with the use of VAs for code analysis and mitigation of vulnerabilities.
The main challenge with this new approach to code analysis is adoption. A recent study involving a small sample of participants shows that currently the primary use of VAs are for music procurement (40\% of users), for information (17\%), and automation (9\%) \cite{bentley2018vas}. Since this is a new avenue of research, there may be initial challenges with adoption in the code analysis arena. However, we believe that as the market grows and coders get exposed to this technology, the adoption rates will increase. Researchers predict a growing use for digital voice assistants over the next few years \cite{tuzovic2018conversational_commerce, klein2020exploringvas}. 

Another challenge with using the PnP model discussed in this research is handling the differences between output reports from different code analyzers. To mitigate this issue, the code analysis community may require standardization of vulnerability reports in popular formats such as XML, JSON, and HTML. Currently, most tools include information such as files, classes, and line numbers where errors are found. While the output formats may be different, NLP techniques such as NER can also be used to mine these reports for key pieces of information to achieve a standard format that can be handled by the virtual assistant and the proposed analysis framework. 

\section{Conclusion}
\label{sec:conclusion}
Getting programmers to write secure code remains a challenge. Security is often sacrificed in an effort to add a feature to a software product or to meet a deadline. When security is sacrificed for other gains, the end result is a product riddled with bugs or vulnerabilities. Steps must be taken to encourage programmers to produce more secure software. In this research, we discuss the limitations of existing code analysis approaches and propose a framework that allows programmers to use virtual assistants to conversationally scan and fix potential vulnerabilities in their code. Virtual assistants are becoming popular in everyday activities such as procuring and listening to music, finding places of interest, managing a smart home, shopping, etc. We posit that as they become more mainstream, they can be used to manage code analysis while keeping programmers productive. We implement our proposed methodology using the Google Assistant and demonstrate its utility in an effort to find new, creative ways to help programmers produce more secure software. Future work will involve extending the model to use any applicable code analyzer based on a plug-and-play paradigm, adding data analytics and visualizations to help programmers draw insights from their code, implementing the refactoring and auto-fixing modules, and conducting a user study to evaluate the framework.


\begin{backmatter}
\section*{Abbreviations}

\begin{abbrvlist}
\item[DAST] Dynamic application security testing
\item[IAST] Interactive application security testing
\item[NLP] Natural language processing
\item[PnP] Plug-and-play
\item[SAST] Static application security testing
\item[SCAD] Synchronous control asynchronous dataflow
\end{abbrvlist}

\begin{authordetails}
	
	
	\author{Fitzroy D. Nembhard$^{1*}$, Marco M. Carvalho$^{1}$}
	\address[1]{Florida Institute of Technology, Melbourne, Florida, USA}
	%
	\address{*Address all correspondence to: fitzroy@ieee.org}
	
	
	
\end{authordetails}


\end{backmatter}

\end{document}